\documentstyle[prl,aps,amssymb,multicol,epsf]{revtex}
\author{Peter Rupp,
Reinhard Richter, and Ingo Rehberg\\
{\it Physikalisches Institut, Experimentalphysik V,
Universit\"at Bayreuth, D-95440 Bayreuth, Germany}}
\date{\today}
\title{Experimental evidence for directed percolation in spatiotemporal intermittency}
\begin{document}
\maketitle

\begin{abstract}
A new experimental system showing a transition to spatiotemporal intermittency is presented. It consists of a ring of hundred oscillating ferrofluidic spikes. The measured critical exponents of the system agree with those obtained from a theoretical model of directed percolation.
\end{abstract}
\bigskip
PACS: 05.45-a, 47.54+r, 47.52

\begin{multicols}{2}

Spatiotemporal Intermittency (STI) is characterized by patches of ordered and disordered states fluctuating stochastically in space and time. It is often a precursor of chaos in spatially extended systems \cite{roots}.
In 1986 Pomeau \cite{pomeau1986} suggested that the onset of chaos via STI might be analogous to directed percolation (DP) \cite{DP}. Such processes are modeled as a probabilistic cellular automaton with two states per site, associated with the laminar and chaotic patches in the case of STI, and predict some universal properties of STI. In particular, the fraction of chaotic domains is expected to grow with a power law $\epsilon^{\beta}$, where $\epsilon$ measures the distance from threshold. The mean laminar length decreases with $\epsilon^{-\nu_s}$, as does the mean laminar time with $\epsilon^{-\nu_t}$, and the critical distribution is determined by $l^{-\mu_s}$ \cite{DPnumber}. 
\end{multicols}
\begin{table}[medium]
\caption{Experiments and results.}
\begin{tabular}{ l  c  l  c  c  l  c  c  c  c }
Authors          & Year & Experiment & Size & $T_0 (s)$ & Geometry & $\beta$ & $\nu_s$ & $\nu_t$ & $\mu_s$ \\
\hline
Ciliberto et al. \cite{ciliberto1992} & 1988 & RB-convection       & 20      & 10   & annular & --           & $0.5$         & -- & $1.9\pm0.1$ \\
Daviaud et al. \cite{daviaud1990,daviaud1992}& 1990 & RB-convection & 40     & 2    & linear  & $0.3\pm0.05$ & $0.5\pm0.05$  & $0.5\pm0.05$ & $1.6\pm0.2$ \\
~~~~~~~~"                             & "    & ~~~~~~~~~~"         & 30      & 2    & annular & --           & $0.5$         & $0.5$ & $1.7\pm0.1$ \\
Michalland et al.\cite{michalland1993}& 1993 & viscous fingering   & 40      & 1.5  & linear  & $0.45\pm0.05$& $0.5$         & -- & $0.63\pm0.02$ \\
Willaime et al. \cite{willaime1993}   & 1993 & line of vortices    & 15      & 5    & linear  & --           & --            & $0.5$ & -- \\
Degen et al. \cite{degen1996}         & 1996 & Taylor-Dean         & 20 (90) & 1.5  & linear  & $1.30\pm0.26$& $\approx0.64$ & $\approx0.73$ & $1.67\pm0.14$\\
Colovas et al. \cite{colovas1997}     & 1997 & Taylor-Couette      & 30 (70) & 0.5  & linear  & --           & --            & -- & --\\
Bottin et al. \cite{bottin1997}       & 1997 & plane Couette       &  --     & --   & linear  & --           & --            & -- & --\\
Vallette et al. \cite{gollub1997}     & 1997 & fluid fronts        & 40      & 0.5  & linear  & --           & --            & -- & --\\  
Jensen (theory) \cite{DPnumber}       & 1999 & directed percolation & --     & --   & --      & $0.276486(8)$      & $1.096854(4)$       & $1.733847(6)$      & $1.748$\\
present paper                         & --   & ferrofluidic spikes & 108     & 0.08 & annular & $0.3\pm0.05$ & $1.2\pm0.1$   & $0.7\pm0.05$ & $1.7\pm0.05$ 
\end{tabular}
\label{table01}
\end{table}
\begin{multicols}{2}

\begin{figure}
\noindent
\begin{minipage}{7cm}
\epsfxsize=7cm
\epsfbox{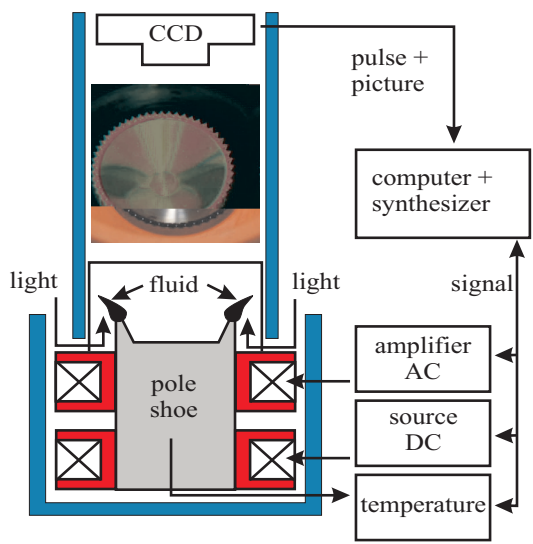}
\caption{Scetch of the experimental setup.}
\label{setup}
\end{minipage}
\end{figure}

Some experimental tests of the conjecture \cite{pomeau1986} in quasi one-dimensional systems have been made \cite{ciliberto1992} -\cite{gollub1997}. A short summary of these and the relevant exponents for the comparison with the DP-model are given in Table\,\ref{table01}. However the statement  "...there is still no experiment where the critical behaviour of DP was seen" by Grassberger \cite{grassberger1997} still seems to be true. Thus in this paper a new experimental approach to this old problem is presented. Motivated by the fact that a single peak of ferrofluid shows chaotic oscillations under external driving of a magnetic field \cite{mahr1998,friedrichs2000}, we introduce here a system where about $100$ of these oscillating peaks are coupled by magnetic and hydrodynamic interaction. They exhibit changes in peak height of about $10$\% and variations in wavelength $\lambda$ of about $50$\%. This system is advantageous because of its short response times and the easy control of the excitation.

The setup of the experiment consists of a cylindrical electromagnet with a sharp edge (Fig.\,\ref{setup}). The magnetic fluid is trapped by the inhomogenous magnetic field at this edge of the magnetically soft iron core. In that way the $40 \,\rm{mm}$ diameter of the pole shoe supports a ring of up to $110$ spikes of magnetic fluid as indicated by the picture in Fig.\,\ref{setup}. The ring of spikes is recorded with a CCD-camera mounted above the pole shoe. The electromagnet contains a bias coil and an excitation coil. The bias coil is provided with a direct current of $I=1.0\,\rm{A}$ to keep the magnetic fluid in its place. This excitation coil is driven by an alternating current, phase-locked with the camera frequency, providing a stroboscopic jitter free recording on long timescales. The fluid is temperature controlled to $12.5\pm0.03\,^{\circ}\rm{C}$ by cooling the pole shoe, and encapsulated in a glass container to provide longterm stability.

\begin{figure}[h]
\noindent
\begin{minipage}{8.6cm}
\epsfxsize=8.6cm
\epsfbox{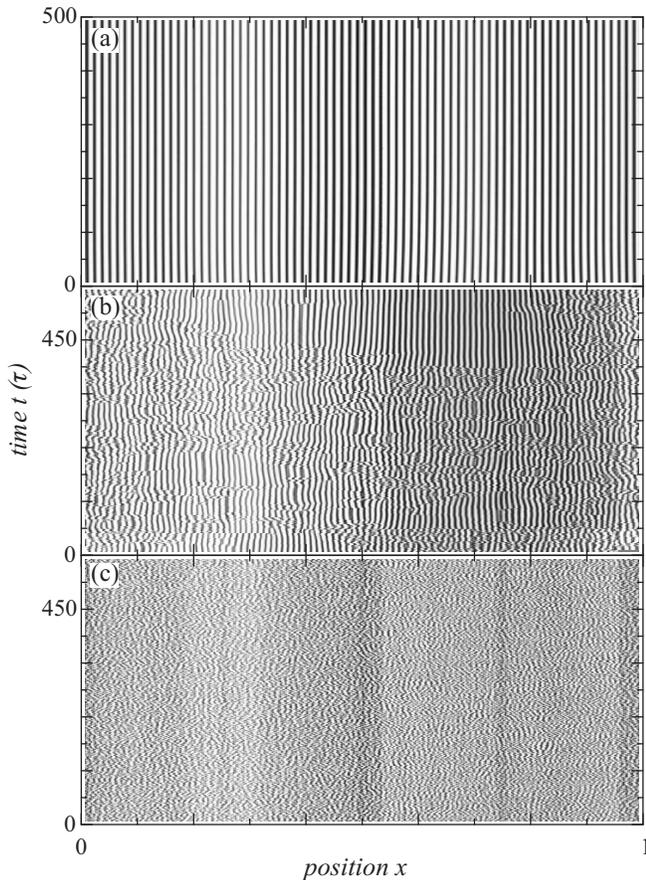}
\caption{Space time plots of the different states of the system. (a) $I_{ex}=2.8\,\rm{A}$: laminar state; (b) $I_{ex}=2.8\,\rm{A}$: spatiotemporal intermittency; (c) $I_{ex}=3.6\,\rm{A}$: chaotic state. $500$ excitation periods are shown. The position is normalized over the size of the ring.}
\label{states}
\end{minipage}
\end{figure}

The spatiotemporal behaviour is investigated by observing the ring with the CCD-camera. To extract the wavelength and amplitude of the spikes as a function of space, we define a ring of interest around the center of the pole shoe which is covering the ring of spikes. It is divided into $1024$ segments. The average of the greyvalues within each segment represents the amplitude. This reduction of the $2D$ image to a single line scan can be done in real time with a frequency of $12.5\,\rm{Hz}$.

\begin{figure}
\noindent
\begin{minipage}{8.6cm}
\epsfxsize=8.6cm
\epsfbox{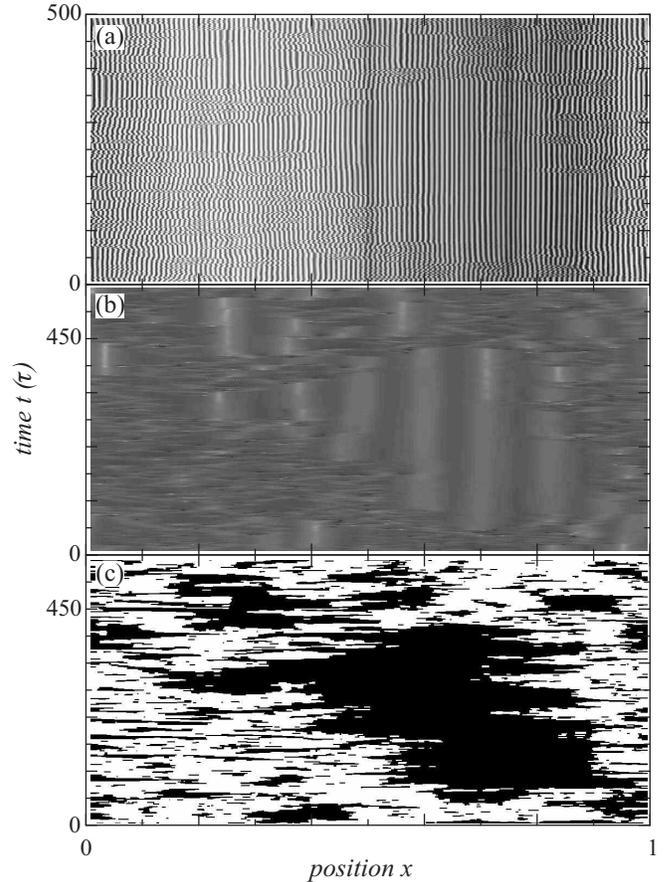}
\caption{x-t-plots at different states of the data processing. (a) raw data; (b) same section as in (a) after transforming the data to complex numbers. White corresponds to large local wavelengths, black to small ones. (c) final step: the relative change of local wavelength over two subsequent periods $\delta\lambda$ is calculated and binarized. The black areas are defined as regular ($\delta\lambda<0.01$), the white are chaotic ($\delta\lambda>0.01$).}
\label{steps}
\end{minipage}
\end{figure}

In Fig.\,\ref{states} $500$ of such scans are shown in a space time plot, where dark regions correspond to high amplitudes. The driving frequency of the excitation field is $f_{ex}=12.5\,\rm{Hz}$, which is the same as the recording frequency. The period $\tau=1/f_{ex}$ is used to scale the time. Due to this stroboscopic recording the regular oscillations of the spikes cannot be seen. At low excitation amplitudes $I_{ex} \ll 3.0\,\rm{A}$ the system is completely regular (Fig.\,\ref{states} (a)) showing $108$ spikes. Slight spatial variations of the wavelength $\lambda$ remain constant in time. In Fig.\,\ref{states} (b) at $I_{ex}=3.0\,\rm{A}$ irregular fluctuations are apparent.

A fairly clear distinction between regular and irregular domains can be made in this image even by naked eye, which we consider as a manifestation of STI. Further increase of $I_{ex}$ leads to spreading of the irregular domains engulfing the regular regions until finally the whole system is chaotic (Fig.\,\ref{states} (c)).

For a clearer distinction between regular and irregular domains (Fig.\,\ref{steps} (a)) we measure the local wavelength $\lambda(x,t)$ \cite{technote}, which is displayed using a grey scale plot in Fig.\,\ref{steps} (b). For the binarization the relative change $\delta\lambda=|\lambda_{t+1}-\lambda_t|/\lambda_t$ of the local wavelength is used: Changes in $\delta\lambda$ which are larger than $0.01$ are counted as irregular, smaller changes belong to regular domains. In Fig.\,\ref{steps} (c) the binarized values of $\delta\lambda$ are indicated. The quantitave analysis of STI described below is based on these binarized information.

\begin{figure}
\noindent
\begin{minipage}{8.6cm}
\epsfxsize=8.6cm
\epsfbox{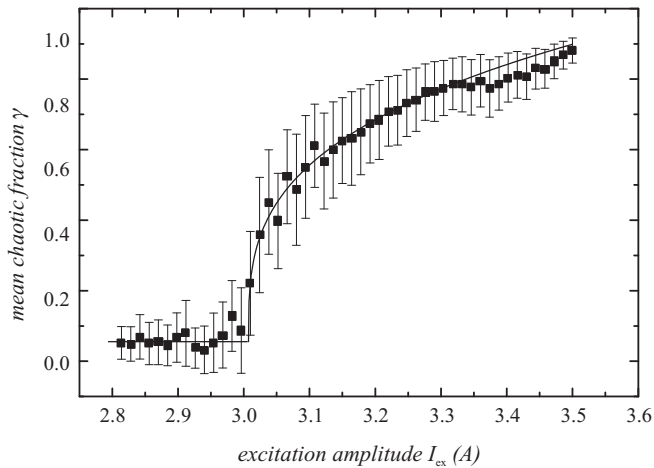}
\caption{The mean chaotic fraction $\gamma$ vs. excitation amplitude $I_{ex}$. The solid line is a power law fit. The errorbars represent the statistical errors of the six measurements.}
\label{cf}
\end{minipage}
\end{figure}

As an order parameter for STI we take the {\it time-averaged chaotic fraction} $\gamma$, which is the ratio of chaotic regions to the whole area of the system. Its variation with the control parameter $I_{ex}$ is shown in Fig.\,\ref{cf}. Here the excitation amplitude is switched from $I_{ex} = 4.05\,\rm{A}$ to the value displayed on the x-axis. $\gamma$ is determined after a waiting time of $\approx 1800\tau$, when the transients have died out. Each data point represents a mean value of six independent runs, which include six refills of the apparatus with fresh fluid.

Close to the onset of STI the mean chaotic fraction is expected to grow with a powerlaw 
$\gamma\sim(I_{ex}-I_c)^{\beta}$.
The solid line in Fig.\,\ref{cf} is a fit to our data, using $I_c$, $\beta$, and an offset representing background noise as adjustable parameters. The threshold value determined in this way is $I_c=3.0\pm0.03\,\rm{A}$ and the exponent $\beta = 0.3\pm0.05$ is in agreement with the theoretical expectation $\beta=0.276486(8)$ \cite{DPnumber}. 

\begin{figure}
\noindent
\begin{minipage}{8.6cm}
\epsfxsize=8.6cm
\epsfbox{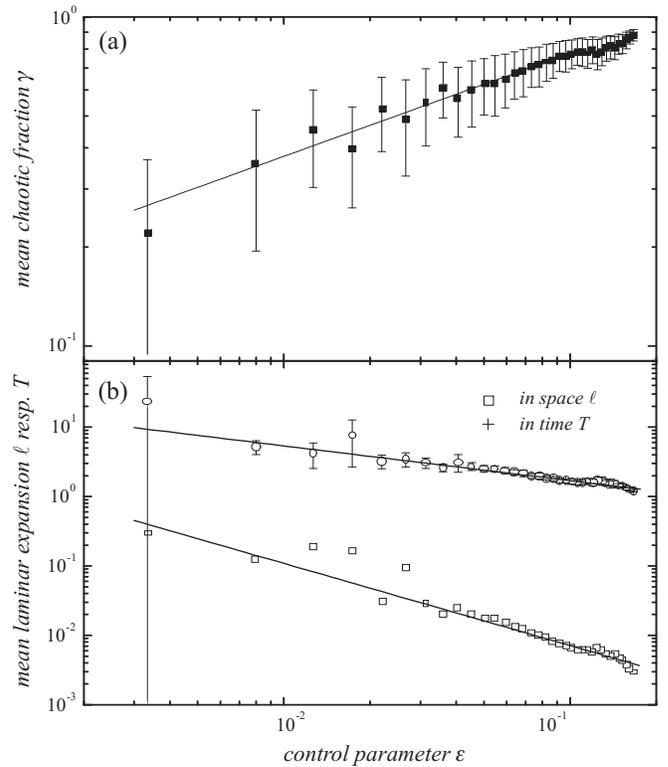}
\caption{(a) The mean chaotic fraction $\gamma$ vs. control parameter $\epsilon$. (b) The mean laminar expansion in space $\ell$ and in time $T$ vs. control parameter $\epsilon$. The solid lines are power law fits.}
\label{cfmll}
\end{minipage}
\end{figure}

An alternative representation of the same data is presented in Fig.\,\ref{cfmll} (a) in a log-log plot. Here the normalized control parameter $\epsilon=I_{ex}/I_c-1$ serves to define the x-axis. In Fig.\,\ref{cfmll} (b) the {\it mean laminar length} $\ell$, which is the average of the spatial expansion of all regular domains at a given $\epsilon$, is presented. It is expected to grow with a power law
$\ell\sim\epsilon^{-\nu_s}$
and the fit yields an exponent $\nu_s=1.2\pm0.1$, which is represented by the solid line. Only data in the range of $0.03<\epsilon<0.1$ are taken into account. For smaller $\epsilon$ a finite size effect is obvious: $\ell$ must be smaller than the system size $1$. For larger $\epsilon$ the system is no longer intermittent, but rather chaotic. The same argument holds for the {\it mean laminar time} $T$, which is the average of the life time of all regular domains (Fig.\,\ref{cfmll}\,(b)). A similar power law behaviour can be seen with an exponent $\nu_t=0.7\pm0.05$. The exponent $\nu_s$ is in good comparison to theoretical expectations of $\nu_s=1.096854(4)$, while $\nu_t$ differs from the theoretical value $\nu_t=1.733847(6)$. 

\begin{figure}
\noindent
\begin{minipage}{8.6cm}
\epsfxsize=8.0cm
\epsfbox{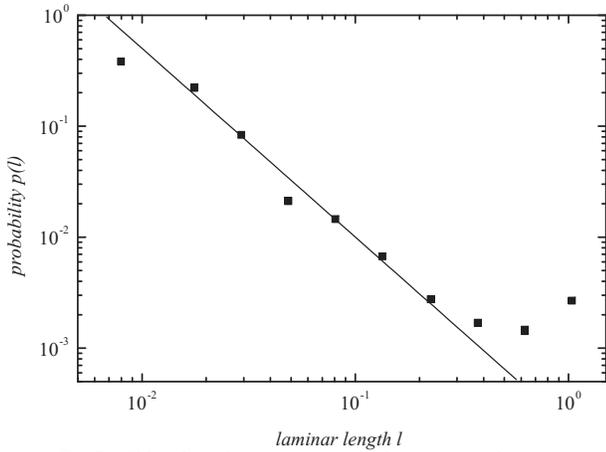}
\caption{Distribution of the laminar domain length $l$ for $\epsilon=0$. The solid line is a power law fit. To suppress the statistical fluctuations the distribution is logarithmical binned.}
\label{distribution01}
\end{minipage}
\end{figure}

In Fig.\,\ref{distribution01} the distribution of the laminar domain length $l$ for $\epsilon=0$ is presented. 
The solid line represents a fit 
$p(l)\sim l^{-\mu_s}$
with $\mu_s=1.7\pm0.05$, in agreement with the theoretical value $\mu_s=1.734$. Within the experimental errors, the theoretically predicted relation
\begin{equation}
\mu_s=2-\frac{\beta}{\nu_s}
\label{eq_relation}
\end{equation}
is thus fulfilled. 

\begin{figure}
\noindent
\begin{minipage}{8.6cm}
\epsfxsize=8.0cm
\epsfbox{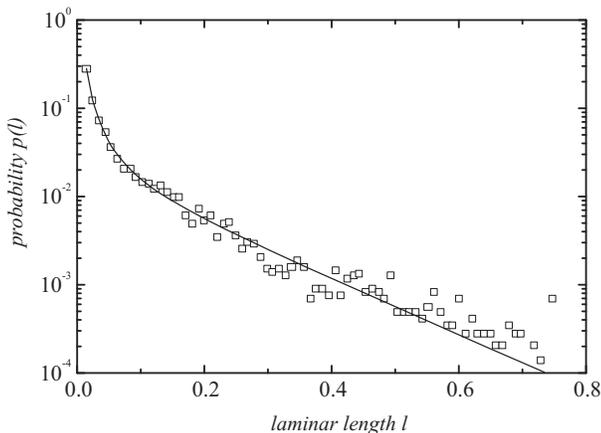}
\caption{Distribution of the laminar domain length $l$ for $\epsilon=0.019$. The solid line is from eq.\,(\ref{eq_powexp}).}
\label{distribution02}
\end{minipage}
\end{figure}

The distribution for $\epsilon>0$ as shown in Fig.\,\ref{distribution02} cannot be described by a simple power law. A more complicated distribution function has been suggested 
\begin{equation}
p(l)=(Al^{-\mu}+B)e^{-l/l_{decay}}
\label{eq_powexp}
\end{equation}
in Ref.\,\cite{ciliberto1992}. The solid line is a fit to this empirical distribution function for $\epsilon=0.019$ with $l_{decay}=0.17$. It shows clearly that the power law is now replaced by a function more reminiscent of an exponential decay.

In summary, we have measured the critical exponents $\beta$, $\nu_s$, $\nu_t$ and $\mu_s$ for the mean chaotic fraction $\gamma$, the mean laminar length $\ell$, the mean laminar time $T$ and the length distribution function. Three of these four exponents agree with the theoretical expectation derived from a DP model and thus fulfill the relation (\ref{eq_relation}). When considering the fact that the theory is applicable only near $I_c$ and that our apparatus is influenced by finite size effects, the results support the idea that our system shows a DP like transition to STI. Further investigations will thus focus on the influence of the system size.

The authors would like to thank Hugues Chat\'e, Haye Hinrichsen and Victor Steinberg for helpful discussions. The experiments have been financially supported by Deutsche Forschungsgemeinschaft through Re588/12.

\end{multicols}

\begin{thebibliography}{16}
\bibitem{technote}
To determine $\lambda$ the scans are transformed to complex values by eliminating amplitudes corresponding to negative wave numbers in Fourier space. $\lambda$ is subsequently calculated as the phase difference between neighbouring values in real space. Details of the algorithm will be presented elsewhere.
\end{thebibliography}
\end{document}